\documentclass[
showpacs,
english,superscriptaddress,nofootinbib,preprintnumbers, aps]{revtex4}
\usepackage{amsmath,amsthm,amssymb,amsfonts}
\usepackage[unicode=true, pdfusetitle, bookmarks=true,bookmarksnumbered=false,bookmarksopen=false, breaklinks=false,pdfborder={0 0 1},backref=false,colorlinks=false]{hyperref}
\usepackage{MnSymbol}
\usepackage{colordvi}
\usepackage{color} 
\usepackage{graphicx}
\usepackage[lofdepth,lotdepth]{subfig}
\usepackage[usenames,dvipsnames]{xcolor}
\usepackage[T1]{fontenc}
\usepackage[latin9]{inputenc}

\newcommand{\m}{\mbox{}}

\newcommand{\be}{\begin{equation}}
\newcommand{\ee}{\end{equation}}
\newcommand{\ba}{\begin{eqnarray}}
\newcommand{\ea}{\end{eqnarray}}

\begin{document}

\title{General relativity in the radial gauge: \\ Reduced phase space and canonical structure}

\author{Norbert Bodendorfer}\email{norbert.bodendorfer@fuw.edu.pl}\affiliation{Faculty of Physics, University of Warsaw, Pasteura 5, 02-093, Warsaw, Poland}
\author{Jerzy Lewandowski}\email{jerzy.lewandowski@fuw.edu.pl}\affiliation{Faculty of Physics, University of Warsaw, Pasteura 5, 02-093, Warsaw, Poland}
\author{J\k{e}drzej \'Swie\.zewski}\email{swiezew@fuw.edu.pl}\affiliation{Faculty of Physics, University of Warsaw, Pasteura 5, 02-093, Warsaw, Poland}
\date{\today}

\begin{abstract}
Firstly, we present a reformulation of the standard canonical approach to spherically symmetric systems in which the radial gauge is imposed. This is done via the gauge unfixing technique, which serves as the exposition in the context of the radial gauge. Secondly, we apply the same techniques to the full theory, without assuming spherical symmetry, resulting in a reduced phase space description of general relativity. The canonical structure of the theory is analyzed.
\end{abstract}

\maketitle

\tableofcontents{}

\section{Introduction}

General relativity viewed from the canonical perspective is a theory governed by constraints. The constraints are first class \cite{HenneauxQuantizationOf} and, therefore, generate gauge transformations. These transformations, being the canonical realization of the general principle of relativity, represent not only the invariance of the physical content of the theory with respect to changes of coordinates, but are also intricately related to the dynamics of the theory. In fact, general relativity is a fully constrained theory, meaning that the Hamiltonian consists only of constraints, which is the central issue in the so-called problem of time. One of the approaches introduced in the literature to deal with the constraints is the procedure of deparametrization \cite{Kijowski, BrownKuchar}. Usually, the procedure makes use of some carefully tailored matter content of the theory to define physical coordinates and, therefore, fixes the freedom in the choice of coordinates. A recent application of deparamterization was discussed in \cite{DuchObservablesFor}, where it has been used in two ways. Firstly, the presence of nonrotating dust was exploited. Deparametrization with respect to it endowed the theory with a preferred notion of time. Such a construction had been known in the literature \cite{BrownKuchar}. The new input of \cite{DuchObservablesFor} was to deparametrize the remaining gauge freedom, of spatial diffeomorphisms, with the use of certain geometrical quantities. The construction was based on the introduction of an observer: a point in the spatial manifold, and a preferred frame which represented the spatial directions as described by the observer. It resulted in a construction of observables invariant with respect to spatial diffeomorphisms (preserving the observer). Moreover, the paper contained a derivation of the Poisson algebra of the observables and some remarks on the possibility of using that construction for solving the diffeomorphism constraint and, hence, obtaining a reduction of the phase space of general relativity. Such a reduction is completed in the current paper.

A related construction of observables has recently been of interest \cite{HeemskerkConstructionOfBulk, KabatDecodingTheHologram, DonnellyDiffeomorphismInvariantObservables}, partially in the context of the AdS/CFT correspondence. The main difference with the construction presented in \cite{DuchObservablesFor} and underlying the current phase space reduction is that in those works, the Hamiltonian constraint is also being gauge-fixed geometrically. By the virtue of the additional gauge condition, the geodesics involved in specifying the observables are spacetime geodesics, as opposed to spatial geodesics in \cite{DuchObservablesFor}.

\section{Motivation}

The construction of the observables in \cite{DuchObservablesFor} can be viewed as expressing the canonical fields in a certain, carefully chosen, geometrical coordinate system. It is a system of coordinates centered in a point which represents the location of the observer. Points of the spatial manifold are then labeled by their proper distance from the observer and the angles which describe the direction in which a given point is located with respect to the observer. The usage of the proper distance from the observer is a very peculiar feature of the construction, and it is transferred to the reduced variables suggested by \cite{DuchObservablesFor}. Namely, it implies a special role played by the surfaces of constant radial distance from the observer. It turns out that the internal geometry of those surfaces (the components of the metric tensor tangent to them) and the tangential components of the momentum of the metric are the variables which suffice (together with fields describing the matter content of the theory, in case they are present) for the description of the geometry. On one hand, this fact enables the reduction of the description of the system to those variables. On the other hand, it seems naturally suited for describing physical systems whose configurations favor a system of concentric structures. Such a situation takes place in the case of spherically symmetric configurations. There the concentric surfaces become spheres, and their internal geometry can be described by a single function related to the area of a given sphere.

The current paper presents a derivation of the reduced phase space, which stems from the insights of \cite{DuchObservablesFor}. The derivation follows a route of gauge fixing, and since spherical coordinates similar to the ones used here have been used many times in the literature, we chose to call the choice of the gauge we are using the {\it radial gauge} (see, e.g., \cite{WilsonMathews}). It leads to reduced variables that have Dirac brackets identical with the corresponding Poisson brackets of the observables constructed in \cite{DuchObservablesFor}. The derivation concerning the full theory is preceded by one performed in the context of spherical symmetry. This serves two purposes. Firstly, it exposes various features of the radial gauge and of the methods that are employed in the current paper, e.g., in the spherically symmetric setting, a certain physical choice is more transparent and the nonlocalities which appear in the Hamiltonian are easier to understand. Secondly, the peculiar feature of the reduced variables of the full theory opened a new possibility for a quantization of the theory. After a translation to connection type of variables and a quantization relying on loop quantum gravity techniques, the geometry becomes effectively $2+1$ dimensional, which is a significant simplification. Because of the peculiar feature of the radial gauge, the quantum theory obtained in this way can readily accommodate a quantum definition of spherically symmetric configurations. In fact, there is more then one definition possessing different advantageous properties. A detailed discussion of the quantization of the reduced phase space obtained in the current paper, together with the identification of the spherically symmetric sector of the theory is presented in a companion paper \cite{BodendorferRadialGaugeII} and summarized in \cite{BodendorferLetter}.

\section{Spherically symmetric midisuperspace}\label{SectionSphericalSymmetry}

In this section, we will apply the radial gauge to the spherically symmetric midisuperspace treatment of general relativity. On one hand, this serves as an introductory exercise to better understand this gauge in the context of full general relativity and to compare our results to previous work in the context of midisuperspaces. On the other hand, we use the resulting Hamiltonian as an ingredient in one of the definitions of spherical symmetry in the companion paper \cite{BodendorferRadialGaugeII} (see also \cite{BodendorferLetter}), since it preserves the respective quantum states as an operator on the full Hilbert space.

\subsection{Canonical structure}

We start following the treatment of the Arnowitt-Deser-Misner (ADM) \cite{ArnowittTheDynamicsOf} formalism for spherical symmetry performed in \cite{KucharGeometrodynamicsOfThe}. The coupling of nonrotating dust is discussed in \cite{VazCanonicalQuantizationOf} (and references therein). We will briefly recall the main steps here.

The gravitational part of the action we will work with is\footnote{Notice that to restore the units convention $G=1=c$ used in \cite{KucharGeometrodynamicsOfThe}, one should put $\chi = 1$ and to restore the units convention commonly used in quantum gravity literature, namely, $8\pi G = 1=c$, one should put $\chi = 8\pi$. Note also that the latter convention will be used throughout Section \ref{sec:FullTheory}.}
\be
	S_{GR}[q] = \frac{\chi}{16\pi} \int dtd^3\sigma N\sqrt{\det q}\left(K^{ij}K_{ij} - (K^i{}_i)^2 + {}^{(3)} R\right).
\ee
The spherically symmetric midisuperspace sector of the ADM formulation of general relativity can be obtained by restricting the spatial line element to
\be
	ds^2 = \Lambda^2(r,t) dr^2 + R^2(r,t) d\Omega^2 ,
\ee
while the restriction affects the lapse function $N(r,t)$ and the shift vector field $\vec{N}(r,t)$ by restricting the shift to only having a radial component $N^r(r,t)$. Both of them are, of course, only functions of $r$ and $t$ in the symmetric context.
The canonical analysis performed in the given variables leads to the following Poisson brackets
\be
\{R(r),\ P_R(\bar{r})\} = \delta(r-\bar{r}), \qquad \{\Lambda(r),\ P_{\Lambda}(\bar{r})\} = \delta(r-\bar{r}),
\ee
where the momenta are defined by
\be\label{DefMomentaSphericalSymmetry}
	P_R = - \frac{\chi \Lambda}{N} \left( \dot R - N^r R' \right) - \frac{\chi R}{N} \left( \dot \Lambda - (\Lambda N^r)' \right),\qquad P_\Lambda = -\frac{\chi R}{N} \left( \dot R - N^r R'  \right),
\ee
where a prime denotes a radial derivative and a dot, a derivative in time.
The spatial diffeomorphism constraint retains only its radial component and reads
\be
	C[\vec N] = \int_0^\infty dr N^r \left(P_R R' - \Lambda P'_\Lambda + C^\text{matt}\right),
\ee
whereas the Hamiltonian constraint is given by 
\be 
	H [N] = \int_0^\infty dr N h = \int_0^\infty dr N  \left(\frac{1}{\chi}\big(\frac{\Lambda P_\Lambda^2}{2 R^2} - \frac{P_R P_\Lambda}{R}\big) + \chi\big(\frac{R R''}{\Lambda} - \frac{R R' \Lambda'}{\Lambda^2} + \frac{R'^2}{2 \Lambda} - \frac{\Lambda}{2}\big) + h^\text{matt}\right),	\label{eq:SphericalHamiltonianConstrainst}
\ee
where $C^\text{matt}$ and $h^\text{matt}$ are contributions from matter fields (in case they are present in the theory).

The addition of nonrotating dust to the theory introduces terms into both of the above constraints; however, after deparametrizing with respect to the dust field (following the prescriptions discussed in \cite{BrownKuchar}), the vector constraint regains the above form, while the Hamiltonian constraint is replaced by a true Hamiltonian equal to
\be
	H[1] = \int_0^\infty dr h.
\ee

\subsection{Mass functional}

As it was noted in \cite{KucharGeometrodynamicsOfThe}, in spherical symmetry, it is possible to give a local definition of a mass functional. One can check by explicit calculation that the expression
\be\label{MassFunctional}
	m := \frac{1}{2}\left(\frac{P_\Lambda^2}{R} + R\left(1 - \frac{{R'}^2}{\Lambda^2}\right)\right)
\ee
has weakly vanishing Poisson brackets with the vacuum vector and Hamiltonian constraints. Moreover, its spatial derivative vanishes on the constraint surface, and therefore, the functional is in fact a constant throughout space and time. The reason why this functional represents the mass becomes apparent once asymptotic conditions and possible boundary terms are taken into account (this is discussed in Appendix \ref{AppAsymptotics}) or when the Schwarzschild solution is investigated (see Appendix \ref{AppSchwarzschild}).

\subsection{Gauge (un)fixing --- measuring distance from zero} \label{SectionGaugeUnFixingSSFromZero}

The next step is to implement the radial gauge. We choose 
\be\label{GaugeConditionSS}
	\Lambda(r)=1,
\ee
since it corresponds to choosing the radial coordinate $r$ to measure the proper distance. In this section, we will discuss a construction which corresponds to having $r$ measuring the distance from the center of symmetry (or zero). In Section \ref{SectionGaugeUnFixingSSFromInf}, we will discuss an alternative construction in which the distance will be measured ``from infinity''. 

A technique equivalent to employing the Dirac bracket (with respect to the gauge fixing \eqref{GaugeConditionSS} and the spatial diffeomorphism constraint), is to use gauge unfixing \cite{MitraGaugeInvariantReformulationAnomalous, AnishettyGaugeInvarianceIn}.
This amounts, in the first step, to modify the momentum $P_\Lambda$ in all constraints except for $C$ by powers of $C$ in such a way that it Poisson commutes with the gauge fixing condition on the gauge fixing surface, thus having equal Dirac and Poisson brackets. In a second step, one can interpret $C$ as a gauge fixing for \eqref{GaugeConditionSS} and drop it. Since \eqref{GaugeConditionSS} is Poisson commuting with the remaining constraints (in this case, only the Hamiltonian constraint) after $P_\Lambda$ has been modified as described above inside the other constraints, \eqref{GaugeConditionSS} becomes a first class constraint. As said before, this is classically equivalent to employing the Dirac bracket or using \eqref{GaugeConditionSS} to gauge fix $C$\footnote{Changing the constraint structure, however, might be advantageous in the quantum theory, which was the original motivation for introducing gauge unfixing in \cite{MitraGaugeInvariantReformulationAnomalous}.}. In this paper however, we will not drop $C$ from the list of constraints, but merely use techniques developed in the context of gauge unfixing \cite{AnishettyGaugeInvarianceIn} to express the Hamiltonian in terms of the reduced variables, see equations \eqref{PLambdaInvariantExtension} and \eqref{eq:SphericalHamiltonianConstraint} (in other words, to compute a gauge invariant extension thereof\footnote{Practically, this means adding a power series in the vector constraint to $P_\Lambda$ and inserting the resulting expression in the Hamiltonian. In the case of general relativity (coupled to matter fields), the series for $P_\Lambda$ terminates after the first term and one obtains a manageable expression for the Hamiltonian.}).

\begin{figure}[h]
\begin{center}\includegraphics[width=0.5\textwidth]{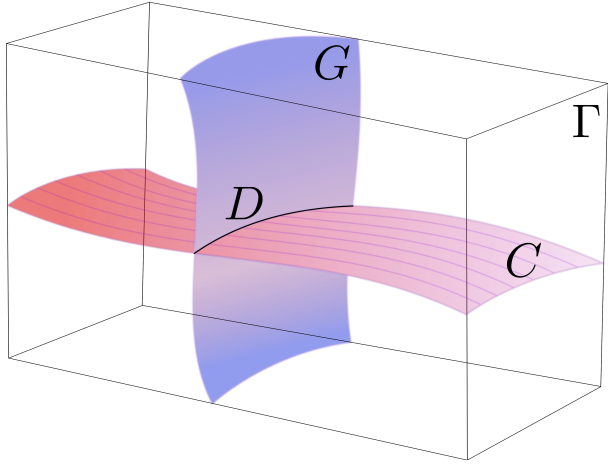}\end{center}
\caption{The picture illustrates schematically the gauge unfixing procedure. $\Gamma$ is a phase space in which the theory is defined. The lined, red (for colors see online) surface $C$ is the constraint surface on which the lines represent gauge orbits. The vertical, blue surface $G$ is the surface on which the gauge fixing condition is satisfied. The thick black line $D=C\cap G$ denotes the points in $\Gamma$ that we are most interested in. The procedure of gauge unfixing provides us with an extension of the Hamiltonian off $C$ such that the Hamiltonian vector field of the extended Hamiltonian is tangent to both $C$ and $G$, and therefore also to $D$.\label{fig1}
In fact, gauge unfixing defines dynamics (equivalent to the one taking place in $C$) on all of $G$; however, it is the restriction to $D$ which allows for a simple link between the two.
}
\end{figure}

There are two main reasons for which we choose to use the gauge unfixing method. Firstly, it can be seen as a method of directly finding the Hamiltonian preserving the gauge, which in some cases (e.g., the nonsymmetric case in this manuscript) leads to a more straightforward way of obtaining such a Hamiltonian. Secondly, it provides a clear phase space picture of the process of implementing the gauge fixing (see Fig.~\ref{fig1}). However, as has been said before, the gauge unfixing method is equivalent to the standard Dirac procedure of implementing the gauge fixing. This can be best seen in sections \ref{SSWhereToMeasureFrom} and \ref{WhereShouldWeFull}, where the standard approach of solving the constraints in a given gauge is discussed. Also, the shift vectors discussed in Appendix \ref{AppShiftVectors} are closely related to the shift vectors needed to invert the Dirac matrix if one chooses to use the Dirac bracket approach.

Consider an expression linear in $P_\Lambda$, that is, the field smeared with a phase-space-independent function $\mu$. Our goal is to find a function $\underset{\mu}{M}(\bar{r})$ such that, denoting
\be\label{TildedExpressionSS}
	\widetilde{\int_0^\infty dr P_\Lambda(r) \mu(r)} = \int_0^\infty dr P_\Lambda(r) \mu(r) - C[\underset{\mu}{M}]
\ee
the tilded expression satisfies
\be\label{ConditionPLambda}
	\{ \Lambda(r),\ \widetilde{\int_0^\infty dr P_\Lambda(r) \mu(r)}\} = 0
\ee
on the gauge fixing surface $\Lambda(r)=1$. This condition yields the equation
\be\label{EquationForMMu}
	\partial_r \underset{\mu}{M}(r) = \mu(r).
\ee
Since $\underset{\mu}{M}$, being a smearing field of the vector constraint in the spherically symmetric context, has to vanish at zero,\footnote{This is the case for regular spacetimes. See Appendix \ref{AppSchwarzschild} for a discussion of an example of a more general setting.} we have
\be
	\underset{\mu}{M}(\bar{r}) = \int_0^{\bar{r}}d\bar{\bar{r}}\mu(\bar{\bar{r}}).
\ee
It will be convenient to have an expression for $\widetilde{P}_\Lambda(\bar{r})$, which we can obtain from the above procedure, choosing $\mu(\bar{\bar{r}}) = \delta(\bar{r}-\bar{\bar{r}})$. It follows that
\be\label{PTildeIsPMinusC}
	\widetilde{P}_\Lambda(\bar{r}) = P_\Lambda(\bar{r}) - \int_{\bar{r}}^\infty d\bar{\bar{r}} C_r(\bar{\bar{r}}).
\ee
Notice that on the gauge fixing surface
\begin{subequations}\label{PLambdaInvariantExtension}
\begin{align}
	\widetilde{P}_\Lambda(\bar{r}) &= P_\Lambda(\bar{r}) - \int_{\bar{r}}^\infty d\bar{\bar{r}} \left(P_R R' - P'_\Lambda + C^\text{matt} \right)(\bar{\bar{r}}) \\
      &= P_\Lambda(\bar{r}) - \int_{\bar{r}}^\infty d\bar{\bar{r}} \left(P_R R' + C^\text{matt}\right)(\bar{\bar{r}}) - P_\Lambda(\bar{r}) \\
      &= - \int_{\bar{r}}^\infty d\bar{\bar{r}} \left(P_R R' + C^\text{matt}\right)(\bar{\bar{r}}),    \label{ExpressionForPLambda}
\end{align}
\end{subequations}
where we used the fact that $P_\Lambda(r) \rightarrow 0$ when $r \rightarrow \infty$.\footnote{This condition is satisfied for spacetimes which are asymptotically flat. See Appendix \ref{AppAsymptotics} for a more detailed discussion of asymptotical flatness. Note, that in the case of the construction from Appendix \ref{AppSchwarzschild}, this condition is not satisfied; however, in the construction presented there, it is also not necessary.} It is not surprising that $P_\Lambda$ drops out from $\widetilde{P}_\Lambda(\bar{r})$, since by the virtue of \eqref{ConditionPLambda} on the gauge fixing slice
\be\label{PLambdaInvariantExtensionVariation}
	\frac{\delta \widetilde{P}_\Lambda}{\delta P_\Lambda} = 0.
\ee

We are now ready to gauge (un)fix our model. This is done by setting $\Lambda(r) = 1$ and using the expression \eqref{ExpressionForPLambda} in the Hamiltonian.
The Hamiltonian turns out to be
\begin{multline}
	\left.H[N]\right|_\text{gauge-fix} \\
	= \int_0^\infty dr N\left(\frac{1}{2 \chi R^2}\left(\int_r^\infty d\bar{r} \left(P_R R' + C^\text{matt}\right)(\bar{r})\right)^2 + \frac{P_R}{\chi R}\int_r^\infty d\bar{r} \left(P_R R' + C^\text{matt}\right)(\bar{r}) \right.\\
	+ \chi\big(R R'' + \frac{R'^2}{2} - \frac{1}{2}\big) + h^\text{matt}\Bigg).	\label{eq:SphericalHamiltonianConstraint}
\end{multline}
It is instructive to look at the procedure of gauge fixing applied above from the perspective of the unreduced phase space. As it is argued in Appendix \ref{AppendixShiftSphericalSymmetryZero}, employing the radial gauge amounts to fixing the shift vector to be
\be
	N_H^r(r) = \frac{1}{\chi}\int_0^r d\bar{r} N(\bar{r})\left(\frac{P_R}{R}(\bar{r}) - \frac{P_\Lambda}{R^2}(\bar{r}) + \frac{1}{2 R^2}(\bar{r})\int_{\bar{r}}^\infty d\bar{\bar{r}} C_r (\bar{\bar{r}})\right).
\ee

In case one couples the model to nonrotating dust and deparametrizes the Hamiltonian constraint with respect to the proper dust time, one obtains the true Hamiltonian in the form
\begin{multline}
	\left.H^\text{true}\right|_\text{gauge-fix} \\
	= \int_0^\infty dr \left(\frac{1}{2 \chi R^2}\left(\int_r^\infty d\bar{r} \left(P_R R'\right)(\bar{r})\right)^2 + \frac{P_R}{\chi R}\int_r^\infty d\bar{r} \left(P_R R'\right)(\bar{r}) + \chi\big(R R'' + \frac{R'^2}{2} - \frac{1}{2}\big)\right). \label{eq:SphericalTrueHamiltonian}
\end{multline}

To sum up, we have solved the spatial diffeomorphism constraint via the gauge unfixing procedure, using the gauge $\Lambda(r)=1$. The reduced phase space turns out to be coordinatized by $R(r)$ and $P_R(r)$, subject to the Hamiltonian constraint \eqref{eq:SphericalHamiltonianConstraint}. Since the Hamiltonian constraint Poisson commutes with the gauge fixing condition due to the addition of spatial diffeomorphism constraints, the Dirac and Poisson brackets are equivalent, and the dynamics can be readily computed using either one. 

Having obtained the reduced phase space, let us compute the equations of motion for the variables in it. For the sake of conciseness, let us spell them out for the case deparametrized with the use of nonrotating dust (with no additional matter) and using the convention $\chi = 1$. They turn out to be
\begin{subequations}\label{EquationsOfMotion}
\begin{align}
	\dot{R}(r) &= -\frac{F(r)}{R(r)} + R'(r)\int_0^r d\bar{r}\left(\frac{P_R(\bar{r})}{R(\bar{r})} - \frac{F(\bar{r})}{R^2(\bar{r})}\right),\\
	\dot{P}_R(r) &= - R''(r) +\frac{P^2_R(r)}{R(r)} - 2\frac{P_R(r)F(r)}{R^2(r)} + \frac{F^2(r)}{R^3(r)} + P_R'(r)\int_0^r d\bar{r} \left(\frac{P_R(\bar{r})}{R(\bar{r})} - \frac{F(\bar{r})}{R^2(\bar{r})}\right),
\end{align}
\end{subequations}
where $F(r)= -\int_r^\infty d\bar{r} \left(P_R R'\right)(\bar{r})$ is the expression \eqref{ExpressionForPLambda} of $\widetilde{P}_\Lambda$ in terms of the reduced variables. These equations are hard to solve in general; however, we can readily identify the trivial solution. Putting $P_R(r)=0$ constantly in time we find
\begin{subequations}
\begin{align}
	\dot{R}(r) &= 0,\\
	0 &= R''(r),
\end{align}
\end{subequations}
which has a unique solution $R(r)=r$ (since by construction $R(0)=0$ and $R'(0)=1$), which is a relation characteristic for flat spacetime. Note that the Hamiltonian vanishes for this solution. As has been demonstrated in \cite{BrownKuchar, HusainPawlowski, Swiezewski}, the Hamiltonian is equal to the momentum of the nonrotating dust field we used for deparametrization. That momentum, in turn, is proportional to the rest mass density of the dust. Therefore, the vanishing of all of those quantities represents a limit in which the dust is not present in the theory, and the spacetime is just the Minkowski spacetime.

\subsection{Gauge (un)fixing --- measuring distance from infinity} \label{SectionGaugeUnFixingSSFromInf}

In this section, we will present a gauge fixing procedure closely related to the one presented in Section \ref{SectionGaugeUnFixingSSFromZero}. The difference here is that we want our gauge fixing to correspond to setting $r$ measuring the proper distance ``from infinity''.

The gauge fixing condition we use in this section is
\be
	\Lambda(r) = 1,
\ee
like it was previously. Finding a gauge invariant extension of $P_\Lambda$ (see equation \eqref{TildedExpressionSS}) amounts to solving the equation (see equation \eqref{EquationForMMu})
\be
	\partial_r \underset{\mu}{M}(r) = \mu(r).
\ee
Instead of integrating it from zero, as we did in the previous section, here we will use the fact that the shift vectors we consider vanish at infinity, which leads to
\be
	\underset{\mu}{M}(\bar{r}) = - \int_{\bar{r}}^\infty d\bar{\bar{r}}\mu(\bar{\bar{r}}).
\ee
This formula leads to the following expression for $P_\Lambda$ as a function of the reduced phase space variables (compare with equation \eqref{ExpressionForPLambda})
\be\label{ExpressionForPLambdaFromInf}
	\widetilde{P}_\Lambda(\bar{r}) = \int_0^{\bar{r}} d\bar{\bar{r}} \left(P_R R' + C^\text{matt}\right)(\bar{\bar{r}}),
\ee
where vanishing of $P_\Lambda$ at zero has been used. The Hamiltonian of the reduced phase space is (compare with equation \eqref{eq:SphericalTrueHamiltonian})
\begin{multline}\label{eq:SphericalHamiltonianConstraintFromInf}
	\left.H[N]\right|_\text{gauge-fix} \\
	= \int_0^\infty dr N\left(\frac{1}{2 \chi R^2}\left(\int_0^r d\bar{r} \left(P_R R' + C^\text{matt}\right)(\bar{r})\right)^2 - \frac{P_R}{\chi R}\int_0^r d\bar{r} \left(P_R R' + C^\text{matt}\right)(\bar{r}) \right.\\
	+ \chi\big(R R'' + \frac{R'^2}{2} - \frac{1}{2}\big) + h^\text{matt}\Bigg).
\end{multline}
It corresponds to fixing the the shift vector to be (derivation is presented in Appendix \ref{AppendixShiftSphericalSymmetryInf})
\be\label{ShiftPreservingGaugeFromInf}
	N_H^r(r) = -\frac{1}{\chi}\int_r^\infty d\bar{r} N\left(\frac{P_R}{R}(\bar{r}) -  \frac{P_\Lambda}{R^2}(\bar{r}) - \frac{1}{2R^2}(\bar{r})\int_0^{\bar{r}} d\bar{\bar{r}}C_r(\bar{\bar{r}})\right).
\ee

\subsection{Where should we measure from?}\label{SSWhereToMeasureFrom}

In this section, we will compare the two alternative variants of implementing the radial gauge, discussing the differences and highlighting the advantages.

The difference between the two ways of ``measuring'' lies in the domains over which certain quantities are integrated. We use the terminology ``measuring from zero'' because the expression for $P_\Lambda(r)$ spelled out in equation \eqref{ExpressionForPLambda} generates transformations of the fields in the reduced phase space in the domain from $r$ to $\infty$. This is what is expected if $r$ measures the distance from zero, since then a variation of the metric at a given $r$ should affect the phase space variables in the radial gauge in points lying further than $r$. This fact becomes straightforward when translated to the language of the observables defined in \cite{DuchObservablesFor}, since the variation of the metric at a given point affects the radial geodesics used in the definition of the coordinates needed for the observables only after they cross the perturbation. Hence, it is the ``measuring from zero'' choice which corresponds more closely to the construction of observables from \cite{DuchObservablesFor}. On the other hand, ``measuring from infinity'' might be more natural in some settings. Firstly, as we will see in Section \ref{SectionFullTheoryFromInfinity}, the option which was suggested in the last section of \cite{DuchObservablesFor} for a solution of the vector constraint corresponds to \eqref{ExpressionForPLambdaFromInf}. Secondly, it naturally accommodates the Schwarzschild solution as discussed in Appendix \ref{AppSchwarzschild}.

It should be noted that the two expressions for $P_\Lambda$ are closely related. The vector constraint in the radial gauge can be rewritten in the form
\be
	P_\Lambda'(r) = \big(P_R R' + C^\text{matt}\big)(r).
\ee
Integrating it from $0$ to $\infty$, using the fact that $P_\Lambda(0)=0$ and assuming\footnote{It is a consequence of the asymptotical flatness conditions discussed in Appendix \ref{AppAsymptotics}. It is is not true, however, in the generalized setting of Appendix \ref{AppSchwarzschild}.} $\lim_{r\rightarrow\infty}P_\Lambda(r)=0$, the following condition is obtained
\be\label{TheCondition}
	\int_0^\infty dr \big(P_R R' + C^\text{matt}\big)(r) = 0.
\ee
This means, that if asymtoptical flatness (in the sense of Appendix \ref{AppAsymptotics}) is assumed, this condition needs to be satisfied throughout the evolution of the system. However, in such a case, the left-hand side of \eqref{TheCondition} Poisson commutes with the Hamiltonian, and therefore, it can be viewed as a restriction on the initial data consistent with the dynamics.

The condition \eqref{TheCondition} renders the two expressions for $P_\Lambda$ numerically equivalent (they necessarily sum up to zero). However, they are distinct when viewed as functionals on the reduced phase space of the theory, since they generate diffeomorphisms of the canonical data in different domains.

\subsection{Summary of the spherically symmetric setting}

In the context of canonical spherically symmetric general relativity, we have presented derivations of two reduced phase spaces. Each is parametrized by two functions, $R(r)$ and $P_R(r)$, of which the first describes the area (given by $4\pi R^2$) of a sphere located at the proper distance $r$ from the center of symmetry, while the latter is its conjugate momentum. Their dynamics is governed by the Hamiltonian constraint \eqref{eq:SphericalHamiltonianConstraint} or \eqref{eq:SphericalHamiltonianConstraintFromInf} (in case we additionally deparametrize with the use of nonrotating dust, by the true Hamiltonian \eqref{eq:SphericalTrueHamiltonian} or its analogue), yielding equations of motion spelled out in \eqref{EquationsOfMotion} or their analogue. The reduced phase spaces support asymptotically flat spacetimes (see section \ref{AppAsymptoticsSS} of Appendix \ref{AppAsymptotics} for details), and the initial data is subject to the condition \eqref{TheCondition}.  One of the two phase spaces may be modified such that the Schwarzschild solution can be readily identified, for details, see Appendix \ref{AppSchwarzschild}.

\section{Full theory} \label{sec:FullTheory}

\subsection{Canonical structure and gauge fixing}

We will now turn to the full theory and rederive the main results of \cite{DuchObservablesFor} via gauge fixing. This way of derivation is complementary to the explicit construction of the observables as performed in \cite{DuchObservablesFor} and leads, as is clear by general arguments \cite{AnishettyGaugeInvarianceIn}, to the same observable algebra. It was the basic idea of \cite{DuchObservablesFor} to define a physical coordinate system by specifying a point $\sigma_0$ in a spatial slice $\Sigma$ of the spacetime manifold and using the exponential map from the tangent space $T_{\sigma_0}\Sigma$ to $\Sigma$ to define such coordinates. Since the exponential map relies on radial geodesics, such a choice of coordinates can be referred to as the radial gauge. The basic concept of using such a radial gauge is, of course, not new; however, the detailed description of the observables and their Poisson algebra has only been worked out recently in \cite{DuchObservablesFor}.

In order to define local (Gau{\ss}) coordinates $(x^I)$, $I=1,2,3$, we need to choose three linearly independent vectors in $T_{\sigma_0}\Sigma$ which transform properly under spatial diffeomorphisms. This can be accomplished as follows: 
at $\sigma_0$, fix a frame ${e_0}^i_I$, where, in general, ${e_0}^i_I {e_0}^j_J \delta^{IJ} \neq q^{ij}$ and $i,j,\ldots$ are spatial tensor indices on $\Sigma$ in some general coordinate system. Given $q_{ij}$, we now construct the frame $e^a_I$ with ${e}^i_I {e}^j_J \delta^{IJ} = q^{ij}$ as
\be
	e^i_I = \sum_{J=1}^3 M_{IJ} {e_0}^i_J \text{,} \label{eq:DefFrame}
\ee	
where $M_{IJ}$ is a lower triangular matrix.\footnote{In fact, $e^i_I$ is the result of the Gram-Schmidt orthonormalization process performed on ${e_0}^i_I$ with respect to $q_{ij}$.} Now, given an infinitesimal diffeomorphism acting on $q^{ij}$, we have $q_{ij} \rightarrow q_{ij} + \mathcal L_N q_{ij}$. The frame $e^i_I$ constructed as above could {\it a priori} transform as $e^i_I \rightarrow \mathcal L_N e^i_I + L_I \m^J e^i_J$ for some antisymmetric matrix $L_{IJ}$. However, it follows from \eqref{eq:DefFrame} that $L_{IJ} = 0$, meaning that the frame $e^i_I$ transforms exclusively under the same spatial diffeomorphism that $q_{ij}$ transforms under. From this, it follows that the map $x^I \mapsto \exp (x^I e^i_I)$ is spatially diffeomorphism invariant for spatial diffeomorphisms $\Psi$ satisfying
\be
	\Psi(\sigma_0) = \sigma_0, \qquad \Psi'(\sigma_0) = M
\ee
for some lower triangular matrix $M$. In coordinates, the second condition translates to $\partial_I N^J(\sigma_0) = e^i_I \partial_i (N^j e_j^J) (\sigma_0) $ on the vector field $N^i$ generating $\Psi$. Such diffeomorphisms will be later referred to by Diff$_\text{obs}$ as they where coined in \cite{DuchObservablesFor}.
These coordinates have only a finite range in general; however, they are always defined in a neighborhood of $\sigma_0$, and their range is maximal in case of spherical symmetry (up to nontrivial topology). In what follows, we assume they are defined globally.

Given the Cartesian ``Gau{\ss}'' coordinates $(x^I)$, we can introduce spherical coordinates $(y^a)=(r, \theta)$. $r$ then labels the geodesic distance from $\sigma_0$, and $\theta$ collectively denotes the two coordinates on surfaces of constant $r$. We will denote the two angular coordinates contained in $\theta$ as $A,B,\ldots$, and $a = r,A$, following the notation of \cite{DuchObservablesFor}. Tensor indices can now also be written in terms of these coordinates. For a general tensor density $T$ of weight $\omega$ and general coordinates $(z^i)$, we can write both
\begin{subequations}
\begin{align}
{T^{IJ\ldots}}_{KL\ldots} &= \left|\frac{\partial(z^1,z^2,z^3)}{\partial (x^1,x^2,x^3)}\right|^\omega\frac{\partial x^I}{\partial z^i}\frac{\partial x^J}{\partial z^j}\ldots \frac{\partial z^k}{\partial x^K}\frac{\partial z^l}{\partial x^L}\ldots {T^{ij\ldots}}_{kl\ldots},\\
{T^{ab\ldots}}_{cd\ldots} &= \left|\frac{\partial(z^1,z^2,z^3)}{\partial (r,\theta^1,\theta^2)}\right|^\omega\frac{\partial y^a}{\partial z^i}\frac{\partial y^b}{\partial z^j}\ldots \frac{\partial z^k}{\partial y^c}\frac{\partial z^l}{\partial y^d}\ldots {T^{ij\ldots}}_{kl\ldots}.
\end{align}
\end{subequations}
We will call the two introduced coordinate systems \emph{adapted to the metric $q_{ij}$}. Note that although both $(x^I)$ and $(y^a)$ depend on the metric $q_{ij}$, the relation between them ($x^I=rn^I$, where $n^I$ is a unit vector field depending only on $\theta$) is independent of the metric.

We will work within the ADM formulation \cite{ArnowittTheDynamicsOf} of general relativity coupled to matter fields, that is with canonical brackets 
\be
	\{q_{ij}(\sigma),\ p^{kl}(\bar{\sigma}) \} = \delta(\sigma, \bar{\sigma}) \delta_{(i}^k \delta_{j)}^l, \qquad \{ \phi_\alpha(\sigma),\ \pi^{\alpha'}(\bar{\sigma}) \} = \delta(\sigma, \bar{\sigma}) \delta_\alpha^{\alpha'} \text{.}
\ee
$i,j,\ldots$ are spatial tensor indices on $\Sigma$ and $\alpha, \alpha'$ label schematically the different matter field species. We restrict here to scalar fields for concreteness. 

Basically, one would now like to introduce the gauge $q_{ra} = \delta_{ra}$, where $r,a$ are coordinates adapted to $q_{ab}$, in order to use the coordinates $a=r,A$ as the physical spherical coordinates described earlier. However, this gauge condition does not impose any restriction on $q_{ij}$, since any metric satisfies this property when expressed in its adapted coordinates.
We therefore specify a reference metric $\check q_{ij}$ on $\Sigma$. This metric induces the spherical $\check a$ and the Cartesian $\check I$ adapted coordinates via the above procedure. By construction, we have $\check q_{\check r \check a}(\sigma) = \delta_{\check r \check a}$. It will now be crucial to observe that given a second metric $q_{ij}$ on $\Sigma$, which in the coordinates adapted to $\check q_{ij}$ reads $q_{\check r \check a} = \delta_{\check r \check a}$, we have $q_{\check a \check b} dy^{\check a} dy^{\check b} = \check q_{\check a \check b} dy^{\check a} dy^{\check b} + \hat q_{\check A \check B} dy^{\check A} dy^{\check B}$ for some $\hat q_{\check A \check B}$. In simple terms, $q_{\check a \check b}$ and $\check q_{\check a \check b}$ differ only in the $\check A \check B$ components, if and only if, they induce the same adapted coordinate system \cite{DuchObservablesFor}. Moreover, if  $q_{\check r \check a} = \delta_{\check r \check a}$, then the coordinates adapted to $q_{ij}$ and $\check q_{ij}$ coincide.

Using the coordinates adapted to $\check q_{ij}$, we can impose the gauge condition $q_{\check r \check a}(\sigma) = \delta_{\check r \check a}$ for the spatial diffeomorphism constraint. Since the ``check'' and ``hat'' were introduced only for clarity of the argument in the previous paragraph, we will now stick to the ``checked'' coordinates dropping the notation with the ``check''. To proceed with the gauge fixing, we need to show that the Poisson bracket of the gauge fixing condition with the vector constraint, namely,
\be
	\left\{ q_{ra}(\sigma),\ C[\vec{N}] \right\} = 2 N_{(r;a)}(\sigma)   \label{PoissonGaugeCond}
\ee
is invertible on the constraint surface. This problem has already been solved in \cite{DuchObservablesFor}, where the equation 
\be
	2 \nabla_{(r} N_{a)} = \omega_{ra}    \label{EquationFromJHEP}
\ee
for arbitrary (symmetric) $\omega_{ra}$ has been solved for the vector field $N^a$ generating an element of Diff$_{\text{obs}}$. The solution reads
\begin{multline}\label{DiffObsN}
 \vec{N}(r,\theta) = \left[ \frac{1}{2} \overline{\omega}_{KJ}(0) h^{JL}rn^K \right] \partial_L
+\frac{1}{2}\left[\int_0^r d\bar{r}\ \omega_{rr}(\bar{r},\theta)\right] \partial_{r}\\
+\left[ \int_0^r d\bar{r}\ q^{BA}(\bar{r},\theta) \left(\omega_{rA}(\bar{r},\theta) - \frac{1}{2}\partial_A \left(\int_0^{\bar{r}}d\bar{\bar{r}}\ \omega_{rr}(\bar{\bar{r}},\theta) \right) \right) \right] \partial_B,
\end{multline}
where $n^I$ is a unit vector field such that $x^I=rn^I$, $h^{IJ}=\delta^{IJ}-n^In^J$ and $\overline{\omega}_{KJ}$ is built from the elements of $\omega_{KJ}$ in the following way\footnote{The bar in $\overline\omega$, being a notation inherited from \cite{DuchObservablesFor}, denotes a certain operation on $\omega$, unlike in $\bar{r}$ and $\bar{\sigma}$, where it is used to denote a point different than $r$ or $\sigma$, respectively.}
\be
 \begin{bmatrix}
  \overline{\omega}_{11} & \overline{\omega}_{12} & \overline{\omega}_{13}\\ 
  \overline{\omega}_{21} & \overline{\omega}_{22} & \overline{\omega}_{23}\\ 
  \overline{\omega}_{31} & \overline{\omega}_{32} & \overline{\omega}_{33}\\ 
 \end{bmatrix}  
 =
  \begin{bmatrix}
  \omega_{11} & 0 & 0\\ 
  \omega_{21}+\omega_{12} & \omega_{22} & 0\\ 
  \omega_{31}+\omega_{13} & \omega_{32}+\omega_{23} & \omega_{33}\\ 
 \end{bmatrix}.
\ee

In the following, it will be convenient to rewrite this solution as\footnote{Note, that the integrals in equations from \eqref{eq:DABDefinition} to \eqref{eq:PoissonPra} are not really geometrically defined, as the notation may suggest. In fact they are defined as integrals in the ``checked'' coordinates mentioned above.}
\be
	N^a_{[\omega]}(\sigma) = \int_\Sigma d^3 \bar{\sigma}  D^{-1 \, ab}(\sigma, \bar{\sigma}) \omega_{rb} (\bar{\sigma}) \label{eq:DABDefinition}
\ee
with $D^{-1 \, ab}(\sigma, \bar{\sigma})$ satisfying 
\be
	\int_\Sigma d^3 \bar{\sigma} D_{ab} (\sigma, \bar{\sigma}) D^{-1 \, bc}(\bar{\sigma}, \bar{\bar{\sigma}}) = \delta_a^c \delta(\sigma, \bar{\bar{\sigma}}) \text{.} \label{eq:DDInverse}
\ee
One more useful notation we introduce is the smeared gauge fixing condition, namely, for a field $\nu^a$
\be
	q_{ra}[\nu^a] := \int_\Sigma d^3\sigma q_{ra}(\sigma)\nu^a(\sigma).
\ee
Note that with the notation we introduced, we can rewrite \eqref{PoissonGaugeCond} in the form
\be
	 \{q_{ra}[\nu^a],\ C[\vec N]\} = \int_\Sigma d^3 \sigma \int_\Sigma d^3 \bar{\sigma} \, \nu^a(\sigma) D_{ab}(\sigma, \bar{\sigma}) N^b(\bar{\sigma}) =: D[\nu,\vec N].
\ee
The Dirac matrix reads
\be
 \{ \left( \begin{array}{c}
C[\vec{M}]  \\
q_{ra}[\mu^a] \end{array} \right),\ \left( \begin{array}{ccc} C[\vec{N}], q_{rb}[\nu^b] \end{array} \right) \} 
=
\left( \begin{array}{cc}
C[\mathcal L_{\vec{M}} \vec{N}] & -D[\nu, \vec M] \\
D[\mu, \vec N] & 0 \end{array} \right)
\ee
and can be easily inverted on the constraint surface $C[\vec{N}_{\text{obs}}] = 0 = q_{ra} - \delta_{ra}$ by using \eqref{eq:DDInverse}. The Dirac bracket between two phase space functions $F$ and $G$ now reads
\begin{align}
	\left\{F,\ G \right\}_{\text{DB}} = \left\{F,\ G \right\}  &- \int_\Sigma d^3 \sigma \int_\Sigma d^3 \bar{\sigma} \{F,\ C_a(\sigma)\} D^{-1 \, ab}(\sigma, \bar{\sigma}) \{  q_{rb}(\bar{\sigma}),\ G\}    \nonumber\\
										&- \int_\Sigma d^3 \sigma \int_\Sigma d^3 \bar{\sigma} \{F,\ q_{ra}(\sigma)\} D^{-1 \, ab}(\sigma, \bar{\sigma}) \{C_b(\bar{\sigma}),\ G\}.
\end{align}
Let us now compute the Dirac brackets between the elementary phase space coordinates $q_{AB}, p^{AB}, q_{ra}, p^{ra}$ to test if we recover the same algebra as the corresponding observables formed in \cite{DuchObservablesFor} (see also remarks on that algebra discussed in \cite{DuchAddendum}). First, the bracket
\be
	\left\{q_{AB}(\sigma),\ p^{CD}(\bar{\sigma}) \right\}_{\text{DB}} = \left\{q_{AB}(\sigma),\ p^{CD}(\bar{\sigma}) \right\} = \delta(\sigma, \bar{\sigma}) \delta_{(A}^C \delta_{B)}^D
\ee
follows directly. Next, we compute
\begin{align}
	\left\{q_{ra}[\nu^a],\ F\right\}_{\text{DB}} 	&=  \left\{q_{ra}[\nu^a],\ F\right\} - \int_\Sigma d^3 \sigma \int_\Sigma d^3 \bar{\sigma} \{q_{ra}[\nu^a],\ C_c(\sigma)\} D^{-1 \, cb}(\sigma, \bar{\sigma}) \{  q_{rb}(\bar{\sigma}),\ F\} \nonumber\\
									&= \left\{q_{ra}[\nu^a],\ F\right\} - \int_\Sigma d^3 \sigma \int_\Sigma d^3 \bar{\sigma} \int_\Sigma d^3 \bar{\bar{\sigma}} \, \nu^a(\bar{\bar{\sigma}}) D_{ac}(\bar{\bar{\sigma}}, \sigma) D^{-1 \, cb}(\sigma, \bar{\sigma}) \{  q_{rb}(\bar{\sigma}),\ F\}\nonumber\\
									&= \left\{q_{ra}[\nu^a],\ F\right\} -  \left\{q_{ra}[\nu^a],\ F\right\} = 0
\end{align}
for an arbitrary $F$, which exemplifies that it is consistent to impose $q_{ra} = \delta_{ra}$ before computing the Dirac bracket. This corresponds to the identity $Q_{ra}(r,\theta) = \delta_{ra}$ for the diffeomorphism invariant observables of \cite{DuchObservablesFor}. The most interesting Poisson brackets involve $p^{ra}$. For $F$ independent of $p^{rb}$, we compute (for clarity, we consider $p^{ra}$ smeared with a field $\kappa_a$)
\begin{align}
	\left\{F,\ p^{ra}[\kappa_a] \right\}_{\text{DB}} &=\left\{F,\ p^{ra}[\kappa_a]\right\}  	- \int_\Sigma d^3 \sigma \int_\Sigma d^3 \bar{\sigma} \{F,\ C_b(\sigma)\} D^{-1 \, bc}(\sigma, \bar{\sigma}) \{q_{rc}(\bar{\sigma}),\ p^{ra}[\kappa_a] \}   \nonumber \\
							&=	\left\{F,\ p^{ra}[\kappa_a]\right\}	- \int_\Sigma d^3 \sigma \int_\Sigma d^3 \bar{\sigma} \{F,\ C_b(\sigma)\} D^{-1 \, bc}(\sigma, \bar{\sigma})\kappa_c(\bar{\sigma})   \nonumber \\
							&=	\left\{F,\ p^{ra}[\kappa_a]\right\} - \mathcal L_{\vec N_{[\kappa]}} F   \label{eq:PoissonPra}
\end{align}
Also, this Dirac bracket agrees with the observable algebra from \cite{DuchObservablesFor}. Finally, the bracket $\left\{p^{ra}[\kappa_a],\ p^{rb}[\lambda_b]\right\}_{\text{DB}}$ follows analogously.

\subsection{Gauge (un)fixing in the full theory --- measuring distance from zero}\label{GaugeFixingFullFromZero}

Complicated Dirac brackets such as \eqref{eq:PoissonPra} are an obstacle to quantization. While the reduced phase space is already coordinatized by the canonical pair $q_{AB}$, $p^{AB}$, as well as the matter fields, the Hamiltonian still contains $p^{ra}$ as a problematic part. Therefore, one either needs a representation of the Dirac algebra containing $p^{ra}$, or we need to express $p^{ra}$ classically in terms of the remaining variables by adding constraints. A means to do this is the formalism of gauge unfixing \cite{MitraGaugeInvariantReformulationAnomalous, AnishettyGaugeInvarianceIn} implemented above in the spherically symmetric case. Here, it boils down to computing a gauge invariant extension of a phase space function canonically conjugate to the variable being gauge fixed, namely $p^{ra}$, with respect to the gauge flow of the gauge fixing condition $q_{ra} - \delta_{ra}=0$, by adding terms proportional to the Diff$_{\text{obs}}$ spatial diffeomorphism constraint. Let us consider a phase space independent, symmetric smearing tensor $\mu_{ab}$, which will later be limited to have only some nonvanishing components. We are interested in
\be\label{DefinitionOfPTilde}
	\widetilde{\int p^{ab}\mu_{ab}} = \int p^{ab}\mu_{ab} - C[\underset{\mu}{\vec M}],
\ee
where the vector constraint $C$ contains both the gravitational part and possibly a contribution from some matter content of the theory (denoted below by $C^{\text{matt}}$). The vector field $\underset{\mu}{\vec M}$ should be chosen such that the condition
\be \label{ConditionOnPTilde}
	\{q_{ra},\ \widetilde{\int p^{ab}\mu_{ab}}\} = 0
\ee
holds on the gauge fixing surface. This condition translates to
\be
	\mu_{ra} = 2\underset{\mu}{M}{}_{(r;a)},
\ee
which is an equation of the type \eqref{EquationFromJHEP} and therefore, we can readily spell out its solution
\begin{multline}\label{VectorFieldForPTilde}
 \underset{\mu}{\vec{M}}(r,\theta) = \left[ \frac{1}{2} \overline{\mu}_{KJ}(0) h^{JL}rn^K \right] \partial_L
+\frac{1}{2}\left[\int_0^r d\bar{r}\ \mu_{rr}(\bar{r},\theta)\right] \partial_{r}\\
+\left[ \int_0^r d\bar{r}\ q^{BA}(\bar{r},\theta) \left(\mu_{rA}(\bar{r},\theta) - \frac{1}{2}\partial_A \left(\int_0^{\bar{r}}d\bar{\bar{r}}\ \mu_{rr}(\bar{\bar{r}},\theta) \right) \right) \right] \partial_B.
\end{multline}
For the time being, let us consider only fields $\mu$ which are vanishing at zero. For such fields, the first term in the above solution does not contribute. Considering first $\mu$ to be such that
\be
	\mu_{rA}(r,\theta) = \delta(r,r_0)\delta(\theta,\theta_0)q_{AA_0}(r,\theta),
\ee
we obtain
\be
	\underset{\mu}{\vec{M}}(r,\theta) = \Theta(r-r_0)\delta(\theta,\theta_0)\partial_{A_0}.
\ee
When implemented in \eqref{DefinitionOfPTilde}, the above vector fields give
\be\label{PTildeRA}
	\widetilde{p}^r{}_{A_0}(r_0,\theta_0) = \int_{r_0}^\infty dr \left(\overset{0}{\cal D}{}_B p^B{}_{A_0}(r,\theta_0) - \frac{1}{2} C^{\text{matt}}_{A_0}(r,\theta_0)\right) + \underset{r\rightarrow\infty}{\lim}p^r{}_{A_0}(r,\theta_0),
\ee
where a notation from \cite{DuchObservablesFor} has been employed, namely, $\cal D$ denotes the derivative covariant with respect to the tangential metric $q_{AB}$, and the zero above the symbol means the derivative is with respect to the $\theta_0$ variable. Until the end of Section \ref{GaugeFixingFullFromZero}, we will drop the last term from the above formula, assuming it is zero. As will be discussed later (see Section \ref{WhereShouldWeFull} and Appendix \ref{AppAsymptoticsFull}), this is problematic.\footnote{In fact, to be mathematically precise, the limit on the right-hand side of equation \eqref{PTildeRA} should be written in front of both terms together, because only together, they have a finite limit.} However, for the sake of clarity of exposition, we adopt this assumption for the time being.
Note that under that assumption, the expression on the right-hand side does not depend on $p^{ra}$, which is consistent with the condition \eqref{ConditionOnPTilde}. 

Secondly, let us consider $\mu$ to be such that only the
\be
	\mu_{rr}(r,\theta) = \delta(r,r_0)\delta(\theta,\theta_0)
\ee
component is nonvanishing. It gives
\be
	\underset{\mu}{\vec{M}}(r,\theta) = \frac{1}{2}\Theta(r-r_0)\delta(\theta,\theta_0)\partial_r - \frac{1}{2}\Theta(r-r_0)\int_{r_0}^r d\bar{r} q^{AB}(\bar{r},\theta)\big(\partial_A\delta(\theta,\theta_0)\big)\partial_B,
\ee
which leads to the following form of \eqref{DefinitionOfPTilde}
\begin{multline}\label{PTildeRR}
	\widetilde{p}^r{}_{r}(r_0,\theta_0) = -\frac{1}{2}\int_{r_0}^\infty dr \left(\big(p^{AB}q_{AB,r}\big)(r,\theta_0) +  C^{\text{matt}}_{r}(r,\theta_0)\right) \\
+ \int_{r_0}^\infty dr \overset{0}{\cal D}{}_A\left(q^{AB}(r,\theta_0)\int_r^\infty d\bar{r} \left(\overset{0}{\cal D}{}_C p^C{}_B(\bar{r},\theta_0) - \frac{1}{2}C^{\text{matt}}_B(\bar{r},\theta_0)\right)\right),
\end{multline}
where an assumption analogues to the one simplifying equation \eqref{PTildeRA} was adopted.
Note, that expressions \eqref{PTildeRA} and \eqref{PTildeRR} realize our aim; namely, they provide expressions for radial components of the momentum $p^{ab}$ in terms of the reduced variables: the purely angular components of the metric and its momentum and possibly the matter fields.

The only part of $\widetilde{p}^r{}_a$ we have not addressed yet is its behavior at zero. However, since ${p}^{ij}$ is a tensor density, the components $p^r{}_a$ vanish at zero. Therefore, we choose $\widetilde{p}^r{}_a$ to also vanish at zero (since it has to do so on the constraint surface anyway). One might worry that in this way, we disregarded the first term in \eqref{VectorFieldForPTilde}, but we will see later that such a choice is dynamically consistent. Actually, to guarantee that indeed $\widetilde{p}^r{}_a$ given by \eqref{PTildeRA} and \eqref{PTildeRR} vanish at zero, we need to impose conditions on the canonical data analogous to the condition \eqref{TheCondition}, namely
\begin{subequations}\label{ConditionOnDataFull}
\begin{align}
	\int_{0}^\infty dr &\left(\overset{0}{\cal D}{}_B p^B{}_{A_0}(r,\theta_0) - \frac{1}{2} C^{\text{matt}}_{A_0}(r,\theta_0)\right) = 0,\\
	 -\frac{1}{2}\int_{0}^\infty dr &\left(\big(p^{AB}q_{AB,r}\big)(r,\theta_0) +  C^{\text{matt}}_{r}(r,\theta_0)\right) \nonumber\\
		&+ \int_{0}^\infty dr \overset{0}{\cal D}{}_A\left(q^{AB}(r,\theta_0)\int_r^\infty d\bar{r} \left(\overset{0}{\cal D}{}_C p^C{}_B(\bar{r},\theta_0) - \frac{1}{2}C^{\text{matt}}_B(\bar{r},\theta_0)\right)\right) = 0.
\end{align}
\end{subequations}
Fortunately, also in the current case, these conditions turn out to be preserved by the dynamics, and hence, are just restrictions of the initial data.

Knowing $\widetilde{p}^r{}_a$, we are now ready to spell out the Hamiltonian for the reduced phase space. It is
\be\label{HamiltonianReducedFullTheory}
	\left.H[N]\right|_{\text{gauge-fix}} = \int drd^2\theta N\left( \frac{2}{\sqrt{\det q}}G - \frac{\sqrt{\det q}}{2} {}^{(3)}\!R + h^\text{matt}\right),
\ee
where
\be\label{HamiltonianReducedFullTheoryLegend}
\begin{cases}
	G = \frac{1}{2}(\widetilde{p}^r{}_r)^2 + 2 q^{AB} \widetilde{p}^r{}_A \widetilde{p}^r{}_B - q_{AB}p^{AB}\widetilde{p}^r{}_r + (q_{AC}q_{BD}-\frac{1}{2}q_{AB}q_{CD})p^{AB}p^{CD}, \\
	{}^{(3)}\!R = {}^{(2)}R - q^{AB}q_{AB,rr} - \frac{3}{4} q^{AB}{}_{,r}q_{AB,r} - \frac{1}{4}(q^{AB}q_{AB,r})^2,\\
	\det q = q_{\theta\theta}q_{\phi\phi} - (q_{\theta\phi})^2.
\end{cases}
\ee
The shift vector that corresponds to this Hamiltonian is presented in Appendix \ref{AppendixShiftFullTheoryFromZero}.

\subsection{Gauge (un)fixing in the full theory --- measuring distance from infinity}\label{SectionFullTheoryFromInfinity}

Also in the full theory case, an alternative implementation of the radial gauge can be considered. The vector constraint in the radial gauge has the form
\begin{subequations}
\begin{align}
\partial_{r}p_{\phantom{r}A}^{r} &= -{\cal D}_{B}p_{\phantom{B}A}^{B} + \frac{1}{2}C^{\rm matt}_{A}, \\
\partial_{r}p_{\phantom{r}r}^{r} &= \frac{1}{2}q_{AB,r}p^{AB}-\partial_{A}p_{\phantom{A}r}^{A} + \frac{1}{2}C^{\rm matt}_{r}.
\end{align}
\end{subequations}
The expressions for the $\widetilde p^{r}{}_{a}$ present in \eqref{PTildeRA} and \eqref{PTildeRR} correspond to integrating the above equations from infinity. As it was pointed out in \cite{DuchObservablesFor}, the components $p^r{}_a$ of the momentum have to vanish at zero due to its tensor density character. Therefore, we can integrate the above equations to obtain (we use the tilde to underline that it is an expression for some of the components of the momentum in terms of the variables of the reduced phase space)
\be\label{PTildeRAFromInfinity}
	\widetilde{p}^r{}_{A_0}(r_0,\theta_0) = \int_{0}^{r_0} dr \left(-\overset{0}{\cal D}{}_B p^B{}_{A_0}(r,\theta_0) + \frac{1}{2} C^{\text{matt}}_{A_0}(r,\theta_0)\right)
\ee
and
\begin{multline}\label{PTildeRRFromInfinity}
	\widetilde{p}^r{}_{r}(r_0,\theta_0) = \frac{1}{2}\int_{0}^{r_0} dr \left(\big(p^{AB}q_{AB,r}\big)(r,\theta_0) +  C^{\text{matt}}_{r}(r,\theta_0)\right) \\
+ \int_{0}^{r_0} dr \overset{0}{\cal D}{}_A\left(q^{AB}(r,\theta_0)\int_0^r d\bar{r} \left(\overset{0}{\cal D}{}_C p^C{}_B(\bar{r},\theta_0) - \frac{1}{2}C^{\text{matt}}_B(\bar{r},\theta_0)\right)\right).
\end{multline}
The Hamiltonian governing the dynamics of the reduced phase space is again given by equations \eqref{HamiltonianReducedFullTheory} and \eqref{HamiltonianReducedFullTheoryLegend}; however, the expressions for $\widetilde p^r{}_a$ should now be taken from the above two formulas. This Hamiltonian corresponds to fixing the shift vector field to the one presented in Appendix \ref{AppendixShiftFullTheoryFromInfinity}.

\subsection{Where should we measure from in the full theory?}\label{WhereShouldWeFull}

In this section, we will compare the two implementations of the radial gauge without assuming spherical symmetry, discussing their problems and advantages.

Taking into account the asymptotic behavior of the fields discussed in Appendix \ref{AppAsymptoticsFull}, we see that the derivation presented in Section \ref{GaugeFixingFullFromZero} runs into problems. In particular, \eqref{PTildeRA} and \eqref{PTildeRR} are not well-defined because the integrands are, in general, not integrable at infinity. Those expressions can be viewed as solutions of the vector constraint
\begin{subequations}
\begin{align}
\partial_{r}p_{\phantom{r}A}^{r} &= -{\cal D}_{B}p_{\phantom{B}A}^{B} + \frac{1}{2}C^{\rm matt}_{A}, \\
\partial_{r}p_{\phantom{r}r}^{r} &= \frac{1}{2}q_{AB,r}p^{AB}-\partial_{A}p_{\phantom{A}r}^{A} + \frac{1}{2}C^{\rm matt}_{r}.
\end{align}
\end{subequations}
Applying the asymptotics from Appendix \ref{AppAsymptoticsFull}, we see that ${\cal D}_{B}p_{\phantom{B}A}^{B}$ can, in general, be finite for $r$ going to infinity and hence, it is not integrable in this limit. There are at least two possible ways to deal with this problem. Firstly, one can try to impose a more stringent fall off behavior of the momentum field. This is problematic because of the necessity of such an imposition to be dynamically consistent, {\it i.e.}, preserved by the Hamiltonian. Secondly, one can try to replace the equations for $p^r{}_a$ with equations for $p^{ra}$. It leads to the following form
\begin{subequations}
\begin{align}
\partial_{r}p^{rA} + q^{AB}q_{BC,r}p^{rC} &= -{\cal D}_{B}p^{BA} + \frac{1}{2}q^{AB}C^{\rm matt}_{B}, \\
\partial_{r}p^{rr} &= \frac{1}{2}q_{AB,r}p^{AB}-\partial_{A}p^{rA} + \frac{1}{2}C^{\rm matt}_{r}.
\end{align}
\end{subequations}
The right-hand side of the first equation is integrable in $r$ at infinity; however, the differential operator on the left-hand side is much more complicated. For practical purposes, this operator would need to be explicitly inverted, so that the dependence of the solution on the reduced phase space variables is known. These problems are the drawbacks of the ``measuring from zero'' implementation of the radial gauge. For some applications of the presented construction, it may be an advantage of this implementation that, as shown in Appendix \ref{AppendixShiftFullTheoryFromZero}, it can be guaranteed that the shift vector $\vec N_H$ belongs to the generators of Diff$_\text{obs}$. Moreover, the problems at infinity may be cured by fixing the canonical data at some finite boundary and implementing the construction in the bounded region only.

The implementation of the radial gauge coined ``measuring from infinity'' is favorable for a few reasons. Firstly, unlike the other implementation, this one is in agreement with the asymptotic flatness requirements as spelled out in Appendix \ref{AppAsymptotics}. Secondly, it is the one which allows for a natural description of a Schwarzschild black hole (see Appendix \ref{AppSchwarzschild}). Note also, that this solution was the one already suggested in \cite{DuchObservablesFor}.

\subsection{Summary of the case without spherical symmetry}

In this section, we have derived two reduced phase spaces for general relativity. Each of them is parametrized by a one parameter ($r$ is the parameter) family of intrinsic geometries of the surfaces of constant radial distance from the center, described by $q_{AB}$, and their momenta $p^{AB}$ (possibly also by matter fields, in case they are present in the theory). Their evolution is generated by the Hamiltonian constraint \eqref{HamiltonianReducedFullTheory} (where $\widetilde p^r{}_a$ are given by \eqref{PTildeRA} and \eqref{PTildeRR} or in the other case, by \eqref{PTildeRAFromInfinity} and \eqref{PTildeRRFromInfinity}). Details concerning imposing asymptotical flatness in those reduced phase spaces are described in Appendix \ref{AppAsymptoticsFull}.

\section{Conclusion and outlook}

In this paper, we have used the radial gauge to construct reduced phase spaces for general relativity with and without assuming spherical symmetry.

The spherically symmetric setting allows for a straightforward treatment, since the tensorial structure is considerably simpler (e.g., intrinsic geometry is described by just two functions, the shift vector is a single function). A drawback of the construction is that it results in a nonlocal Hamiltonian, leading to nonlocal equations of motion. On the other hand, the construction for the spherically symmetric case exposes various nontrivial features of the idea and therefore may serve as a toy model for the full case. More importantly, it suits very well a certain quantization scheme (discussed in \cite{BodendorferRadialGaugeII} and \cite{BodendorferLetter}), where it plays an important role in the definition(s) of the spherically symmetric sector of the theory. This is of particular interest, since up to date, the treatments of the simplest (spherically symmetric) collapse scenarios, including quantum gravity effects, have only been performed in midisuperspace models.

The reduction of the nonsymmetric case leads to the same kind of nonlocalities in the Hamiltonian. However, the reduced data still possess a clear geometrical interpretation and allow for a quantization of the system. Furthermore, the splitting of the description into surfaces of constant radial distance from the center and the radial direction is reflected in the quantum theory in a remarkable simplification. The geometry is effectively described by a one parameter family of $2+1$ dimensional geometries, opening the possibility of formulating computable models of quantum dynamics.

There are a few ways in which the presented formalism can be developed further or applied:
\begin{enumerate}
\item The reduced phase spaces suit well a quantization using the loop quantum gravity techniques. Analysis of the symmetric and nonsymmetric cases leads to a definition of a reduction to spherical symmetry on the quantum level. In fact, different definitions of spherical symmetry can be given, retaining different sets of degrees of freedom of the full theory. For the details, we refer the reader to \cite{BodendorferRadialGaugeII} and \cite{BodendorferLetter}.

\item An advantage of the spherically symmetric setting is that radial geodesics spanned from the central point never cross each other. Generically, this does not happen if space is not symmetric. Therefore, in general, the construction we introduced in the nonsymmetric case will break down due to the formation of caustics of the radial geodesics. We excluded this by demanding that the coordinate system we use is global. In general, this issue may also be addressed by modifying our construction, so that we require the radial gauge condition to hold only in some neighborhood of the central point. To be able to consistently perform this modification one would need a (delicate) introduction of a boundary up to which the gauge condition holds.

\item An interesting question to ask is whether the current construction can be generalized to incorporate spatially compact settings. It seems that at least the spherically symmetric case is open for such a generalization, since the point antipodal to the center of symmetry will again be a center of spherical symmetry.

\end{enumerate}

We leave the development of the ideas presented in the second and third points for future research.

\appendix

\section{Shift vector fields corresponding to the chosen gauges}\label{AppShiftVectors}

Our aim in this appendix is to compute the specific shift vector fields, we call them $\vec{N}_H$, which, when included in the Hamiltonian, correspond to the radial gauge in the variants presented above.

\subsection{Spherically symmetric case --- measuring distance from zero}\label{AppendixShiftSphericalSymmetryZero}

The Hamiltonian given by \eqref{eq:SphericalHamiltonianConstraint} is the Hamiltonian of the reduced phase space obtained after fixing the gauge in the variant in which we ``measure from zero''. The question we address in this section is: What is the shift vector field defined in the context of the unreduced phase space such that the Hamiltonian from that phase space (including the vector constraint smeared with the shift we seek) preserves the gauge we chose? In other words, we want to find $\vec{N}_H$ such that
\be
	\left.H[N]\right|_\text{gauge-fix} = H[N] + C[\vec{N}_H].
\ee
This can be easily done by substituting \eqref{PTildeIsPMinusC} instead of \eqref{ExpressionForPLambda} into the Hamiltonian constraint and separating the appropriate terms. After a few manipulations, we can read off
\be
	N_H^r(r) = \frac{1}{\chi}\int_0^r d\bar{r} N\left(\frac{P_R}{R}(\bar{r}) - \frac{P_\Lambda}{R^2}(\bar{r}) + \frac{1}{2 R^2}(\bar{r})\int_{\bar{r}}^\infty d\bar{\bar{r}} C_r(\bar{\bar{r}})\right).
\ee
Note, that this shift vanishes at zero as it should for regular spacetimes in the context of spherical symmetry.

\subsection{Spherically symmetric case --- measuring distance from infinity}\label{AppendixShiftSphericalSymmetryInf}

The shift vector which corresponds to the Hamiltonian given by \eqref{eq:SphericalHamiltonianConstraintFromInf} can be obtained in the same manner yielding
\be
	N_H^r(r) = -\frac{1}{\chi}\int_r^\infty d\bar{r} N\left(\frac{P_R}{R}(\bar{r}) -  \frac{P_\Lambda}{R^2}(\bar{r}) - \frac{1}{2R^2}(\bar{r})\int_0^{\bar{r}} d\bar{\bar{r}}C_r(\bar{\bar{r}})\right).
\ee
Note that the above shift is not automatically vanishing at zero. This fact will be exploited in Appendix \ref{AppSchwarzschild}.

\subsection{Full theory case --- measuring from zero}\label{AppendixShiftFullTheoryFromZero}

In this section, we will discuss the shift vector field which corresponds to fixing the radial gauge in the context of the full theory. The strategy is the same as we used in the context of spherical symmetry. As a starting point, we use
\be
	\tilde{p}^r{}_a(r_0,\theta_0) = p^r{}_a(r_0,\theta_0) - C[\underset{a(r_0,\theta_0)}{\vec M}],
\ee
where
\be
\begin{cases}
	\underset{A_0(r_0,\theta_0)}{\vec M}(r,\theta) = \Theta(r-r_0)\delta(\theta,\theta_0)\partial_{A_0}, \\
	\underset{r(r_0,\theta_0)}{\vec M}(r,\theta) = \frac{1}{2}\Theta(r-r_0)\delta(\theta,\theta_0)\partial_{r} - \frac{1}{2}\Theta(r-r_0)\int_{r_0}^r d\bar{r} q^{AB}(\bar{r},\theta)\left(\partial_A\delta(\theta,\theta_0)\right)\partial_B,
\end{cases}
\ee
in the definition \eqref{HamiltonianReducedFullTheory} of the Hamiltonian. Then separating the Hamiltonian constraint part and extracting the vector constraint part, we can read off the relevant shift vector field $\vec{N}_H$, such that
\be
	\left.H[N]\right|_\text{gauge-fix} = H[N] + C[\vec{N}_H].
\ee
It turns out that
\begin{subequations}\label{NHForFullTheory}
\begin{align}
	N^r_H(r,\theta) = & \frac{1}{2}\int_0^r dr_0 \left(\frac{N}{\sqrt{\det q}}(-p^{rr} + q_{AB}p^{AB} + \frac{1}{2}t_r)\right)(r_0,\theta)\\
	N^A_H(r,\theta) = & \int_0^r dr_0\left(q^{AB}\frac{N}{\sqrt{\det q}}(-4p^r{}_B + 2t_B)\right)(r_0,\theta)\\ 
				& - \frac{1}{2}\int_0^r dr_0 q^{AB}(r_0,\theta)\partial_B \int_0^{r_0} d\bar{r}\left(\frac{N}{\sqrt{\det q}}(-p^{rr} + q_{CD}p^{CD} + \frac{1}{2}t_r)\right)(\bar{r},\theta),
\end{align}
\end{subequations}
where we used the shorthand notation
\be
	t_a(r_0,\theta_0) := C[\underset{a(r_0,\theta_0)}{\vec{M}}].
\ee

A thing to notice is that the above field is almost of the kind spelled out in \eqref{DiffObsN} if one identifies
\begin{subequations}
\begin{align}
\omega_{rr} &= \frac{N}{\sqrt{\det q}}(-p^{rr} + q_{AB}p^{AB} + \frac{1}{2}t_r), \\
\omega_{rA} &= \frac{N}{\sqrt{\det q}}(-4p^r{}_A + 2t_A).
\end{align}
\end{subequations}
As it was pointed out in \cite{DuchObservablesFor}, the radial gauge we are using does not fix the spatial diffeomorphism gauge completely. In fact, it fixes diffeomorphisms called Diff$_{\text{obs}}$, leaving a finite (six) dimensional, residual diffeomorphism freedom. Therefore, it may be desirable to stay within the Diff$_{\text{obs}}$ class when reducing the phase space. The part missing from \eqref{NHForFullTheory} is the contribution from $\omega$ at zero (compare with equation \eqref{DiffObsN}). However, it turns out that we can consistently fix that problem. Notice, that the required contribution from $\omega$ at zero vanishes when we are in a point of the constraint surface in which
\be\label{ConditionInZero}
	\underset{r\rightarrow 0}{\lim}\left(\frac{N}{\sqrt{\det q}}(-p^{rr} + q_{AB}p^{AB})\right) = 0.
\ee
Let us postpone the discussion of the meaning of that condition for now, to notice that the Poisson bracket of it with the gauge fixed Hamiltonian is again proportional to such a term. This means that if we require our initial data to satisfy that condition, the evolution we have formulated will preserve it. To understand the geometrical meaning of the condition we just imposed, let us rewrite it in terms of the extrinsic curvature
\be
	\underset{r\rightarrow 0}{\lim}\left(\frac{N}{2\sqrt{\det q}}(-p^{rr} + q_{AB}p^{AB})\right) = \left(NK_{rr}\right)(0).
\ee
Requiring that it vanishes is equivalent to requiring
\be
	\left(NK_{IJ}\right)(0) = 0 \qquad \forall_{I,J},
\ee
but that means
\be
	-\frac{1}{2}\dot{q}_{IJ}(0) + \partial_{(I}N^H_{J)}(0) = 0.
\ee
Since we know that in the coordinates $I$ we have $q_{IJ}(0) = \delta_{IJ}$ constantly in time, we obtain a condition
\be
	\partial_{(I}N^H_{J)}(0) = 0.
\ee
This condition can easily be interpreted in the light of \cite{DuchObservablesFor}. It just means that from the initial data allowed by the construction without the condition \eqref{ConditionInZero}, we chose such that no relative deformations of the directions associated with the central observer take place as the system evolves, or equivalently, such that the central point (and its infinitesimal neighborhood) is flatly embedded in the spacetime generated by the evolution.

The remaining task is to express the condition \eqref{ConditionInZero} in the reduced phase space variables. In the reduced phase space, we have
\begin{multline}
	p^{rr}(r_0,\theta_0) = -\frac{1}{2}\int_{r_0}^\infty dr \left(\big(p^{AB}q_{AB,r}\big)(r,\theta_0) +  C^{\text{matt}}_{r}(r,\theta_0)\right) \\
+ \int_{r_0}^\infty dr \overset{0}{\cal D}{}_A\left(q^{AB}(r,\theta_0)\int_r^\infty d\bar{r} \left(\overset{0}{\cal D}{}_C p^C{}_B(\bar{r},\theta_0) - \frac{1}{2}C^{\text{matt}}_B(\bar{r},\theta_0)\right)\right),
\end{multline}
so we can rewrite the limit
\begin{multline}
	\underset{r\rightarrow0}{\lim}\left(\frac{N}{\sqrt{\det q}}(-p^{rr}+p^{AB}q_{AB})\right) =\\
	= N(0)\underset{r\rightarrow0}{\lim}\frac{p^{AB}q_{AB} + \frac{1}{2}\int_{r}^\infty \left(p^{AB}q_{AB,r} +  C^{\text{matt}}_{r}\right) - \int_{r}^\infty d\bar{r} {\cal D}_A\left(q^{AB}(\bar{r})\int_{\bar{r}}^\infty \left({\cal D}_C p^C{}_B - \frac{1}{2}C^{\text{matt}}_B\right)\right)}{\sqrt{\det q}}.
\end{multline}
Both the nominator and the denominator vanish as $r\rightarrow0$, so using the L'H\^opital's rule we get
\begin{multline}
	\underset{r\rightarrow0}{\lim}\left(\frac{N}{\sqrt{\det q}}(-p^{rr}+p^{AB}q_{AB})\right) \overset{\text{H}}{=}\\
	= N(0)\underset{r\rightarrow0}{\lim}\frac{\big(p^{AB}q_{AB}\big)_{,r} - \frac{1}{2}p^{AB}q_{AB,r} -  \frac{1}{2}C^{\text{matt}}_{r} + {\cal D}_A\left(q^{AB}\int_{r}^\infty \left({\cal D}_C p^C{}_B - \frac{1}{2}C^{\text{matt}}_B\right)\right)}{\big(\sqrt{\det q}\big)_{,r}}.
\end{multline}
Now we need to look at each of the terms separately
\begin{subequations}
\begin{align}
p^{AB}q_{AB} &\sim r^2 \quad \text{so}\quad \big(p^{AB}q_{AB}\big)_{,r} \sim r,\\
p^{AB}q_{AB,r} &\sim r ,\\
C^{\text{matt}}_{r} &\sim r^2 ,\\
{\cal D}_C p^C{}_B - \frac{1}{2}C^{\text{matt}}_B &\sim r^3 \quad \text{so}\quad {\cal D}_A\left(q^{AB}\int_{r}^\infty \left({\cal D}_C p^C{}_B - \frac{1}{2}C^{\text{matt}}_B\right)\right) \sim r^2 ,\\
\sqrt{\det q} &\sim r^2 \quad \text{so}\quad \big(\sqrt{\det q}\big)_{,r} \sim r ,
\end{align}
\end{subequations}
therefore, using the L'H\^opital's rule again, we get
\be
	\underset{r\rightarrow0}{\lim}\left(\frac{N}{\sqrt{\det q}}(-p^{rr}+p^{AB}q_{AB})\right) \overset{\text{HH}}{=}
	N(0)\underset{r\rightarrow0}{\lim}\frac{\big(p^{AB}q_{AB}\big)_{,rr} - \frac{1}{2}\big(p^{AB}q_{AB,r}\big)_{,r}}{\big(\sqrt{\det q}\big)_{,rr}}
\ee
and the right-hand side is an expression involving only the reduced phase space variables.

\subsection{Full theory case --- measuring from infinity}\label{AppendixShiftFullTheoryFromInfinity}

In this section, we present the shift vector $\vec{N}_H$ which corresponds to the Hamiltonian from Section \ref{SectionFullTheoryFromInfinity}. Rewriting the expressions from \eqref{PTildeRAFromInfinity} and \eqref{PTildeRRFromInfinity} in the form
\be
	\widetilde p^r{}_a = p^r{}_a - C[\underset{a}{\vec M}],
\ee
we can express the reduced Hamiltonian as
\be
	\widetilde H = H + C[\vec N_H],
\ee
with
\begin{subequations}\label{NHForFullTheoryFromInfinity}
\begin{align}
	N^r_H(r,\theta) = & -\frac{1}{2}\int_r^\infty dr_0 \left(\frac{N}{\sqrt{\det q}}(-p^{rr} + q_{AB}p^{AB} + \frac{1}{2}t_r)\right)(r_0,\theta)\\
	N^A_H(r,\theta) = & -\int_r^\infty dr_0\left(q^{AB}\frac{N}{\sqrt{\det q}}(-4p^r{}_B + 2t_B)\right)(r_0,\theta)\\ 
				& - \frac{1}{2}\int_r^\infty dr_0 q^{AB}(r_0,\theta)\partial_B \int_{r_0}^\infty d\bar{r}\left(\frac{N}{\sqrt{\det q}}(-p^{rr} + q_{CD}p^{CD} + \frac{1}{2}t_r)\right)(\bar{r},\theta),
\end{align}
\end{subequations}
where $t_a$ are terms proportional to the vector constraint. This shift vector vanishes at infinity; however, its behavior at zero is more obscure.

\section{Asymptotic conditions on canonical data}\label{AppAsymptotics}

In this appendix we will discuss the falloffs of the canonical fields for which the treatment presented in the main part of the paper is well-defined. We will restrict the discussion to the gravitational sector of the phase space, since it is the nontrivial one. Our analysis will be based on the treatment of asymptotical flatness presented in \cite{BeigMurchadha}.  We demand that there exist asymptotically Minkowskian coordinates $(x^i)$ (additionally, we define a coordinate $\rho:=\sqrt{x^i x^j \delta_{ij}}$ and also, $n^i:=\frac{x^i}{\rho}$) in which the fields behave as
\begin{align}
	q_{ij}(x^k) &= \delta_{ij} + \frac{1}{\rho}s_{ij}(n^k) + O^\infty(\rho^{-1-\epsilon}),\label{AsymptoticsOfMetric}\\
	p^{ij}(x^k) &= \frac{1}{\rho^2}t^{ij}(n^k) + O^\infty(\rho^{-2-\epsilon}),\label{AsymptoticsOfMomentum}
\end{align}
where, as the notation suggests, $s_{ij}$ and $t^{ij}$ are functions of the angles only, and they satisfy the so-called parity properties
\begin{align}
	s_{ij}(n^k) &= s_{ij}(-n^k),\\
	t^{ij}(n^k) &= - t^{ij}(-n^k),
\end{align}
while by $O^\infty(\rho^{-n})$, we denote terms which vanish at infinity as $\rho^{-n}$, their first derivatives vanish as $\rho^{-(n+1)}$, and so on. The asymptotical behavior of the lapse and shift is then
\begin{align}
	N(x^k) &= k(n^k) + O^\infty(\rho^{-\epsilon}),\\
	N^I(x^i) &= k^i(n^k) + O^\infty(\rho^{-\epsilon}),
\end{align}
where
\begin{align}
	k(n^k) &= - k(-n^k),\\
	k^i(n^k) &= - k^i(-n^k).\label{OddShiftAsymptotics}
\end{align}
For such fields, one can check that all the integrals in the canonical theory are convergent ``at infinity" and functionally differentiable. As was argued in \cite{BeigMurchadha}, dropping the parity properties of the lapse and shift gives rise to the ADM four momentum. Here, we loosen the parity property of the lapse only; therefore, the asymptotics of lapse functions we consider are
\be
	N(x^k) = 1 + k(n^k) + O^\infty(\rho^{-\epsilon}).
\ee

\subsection{Spherically symmetric case}\label{AppAsymptoticsSS}

The translation of the asymptotic conditions spelled out in the preceding paragraph to the variables used in spherical symmetry was carried out in \cite{KucharGeometrodynamicsOfThe}. It gave
\begin{subequations}\label{AsymptoticConditionsRho}
\begin{align}
	\Lambda(\rho) &= 1 + \frac{m}{\rho} + O^\infty(\rho^{-1-\epsilon}),\label{AsymptoticConditionsRhoLambda}\\
	R(\rho) &= \rho + O^\infty(\rho^{-\epsilon}),\label{AsymptoticConditionsROfRho}\\
	P_\Lambda(\rho) &= O^\infty(\rho^{-\epsilon}),\\
	P_R(\rho) &= O^\infty(\rho^{-1-\epsilon}),\\
	N(\rho) &= 1 + O^\infty(\rho^{-\epsilon}),\\
	N^r(\rho) &= O^\infty(\rho^{-\epsilon}),
\end{align}
\end{subequations}
where it is easy to check that the formula \eqref{MassFunctional} for the mass functional, yields $m$ in the limit of large $\rho$ when the above asymptotics are employed. This justifies the usage of the same symbol in the definition of the functional and in the asymptotic considerations. It should be noted that the shift may be allowed to be finite at infinity, since it corresponds to an asymptotically odd shift in agreement with \eqref{OddShiftAsymptotics}.

Such conditions require a boundary term to be added to the Hamiltonian constraint to secure its functional differentiability. That boundary term is the ADM energy that in this context is given by
\be\label{ADMEnergy}
	E = \underset{\rho\rightarrow\infty}{\lim}\frac{1}{2}\rho(\Lambda^2 + \frac{R^2}{\rho^2} - \frac{2RR'}{\rho}) = \underset{\rho\rightarrow\infty}{\lim}\frac{1}{2}\rho(\Lambda^2 - 1) = m.
\ee

Note that since the beginning of the discussion of asymptotical flatness we used the variable $\rho$ to denote the radial coordinate denoted by $r$ in the original papers (\cite{BeigMurchadha, KucharGeometrodynamicsOfThe}). This is because in the current paper, the label $r$ is reserved for a specific radial coordinate, namely, the one that measures the proper spatial distance. Since in our analysis we want to use the variable $r$, we need to transform the conditions \eqref{AsymptoticConditionsRho}. What is the relation between $r$ and $\rho$? Making use of the fact that $\Lambda$ is a density and knowing on one hand that it satisfies \eqref{AsymptoticConditionsRhoLambda} and on the other that it is equal to $1$ as a function of $r$, we find the condition
\be\label{AsymptoticConditionLambdaProper}
	\Lambda(\rho)d\rho = 1 \cdot dr.
\ee
After integrating this condition and finding the leading terms in the expansion of $\rho$ as a function of $r$, we find
\be
	\rho(r) = r - m\log r + c + O^\infty(r^{-\epsilon}),
\ee
where $c$ is a constant whose role is to fix the specific value of $\rho$ at a given $r$. For simplicity, we choose it to be zero in the following.
Using this relation in \eqref{AsymptoticConditionsRho} yields the asymptotic conditions on the fields, which we use in the current paper,
\begin{subequations}\label{AsymptoticConditionsR}
\begin{align}
	\Lambda(r) &= 1,\label{AsymptoticConditionsRLambda}\\
	R(r) &= r - m\log r + O^\infty(r^{-\epsilon}),\label{AsymptoticConditionsROfR}\\
	P_\Lambda(r) &= O^\infty(r^{-\epsilon}),\\
	P_R(r) &= O^\infty(r^{-1-\epsilon}),\\
	N(r) &= 1 + O^\infty(r^{-\epsilon}),\\
	N^r(r) &= O^\infty(r^{-\epsilon}).
\end{align}
\end{subequations}
Note, that condition \eqref{AsymptoticConditionsRLambda} is satisfied not only in the limit of large $r$, but for all its values.

Having defined the asymptotic behavior of the fields, we should address the problem of boundary terms. It turns out that the vacuum Hamiltonian and vector constraints are both well-defined (meaning integrable at infinity). However, to ensure the functional differentiability of the Hamiltonian constraint we have to add a boundary term analogous to \eqref{ADMEnergy} with
\be\label{AppADMbounadry term}
	E = \underset{r\rightarrow\infty}{\lim}\frac{1}{2}r(1 + \frac{R^2}{r^2} - \frac{2RR'}{r}) = m.
\ee

\subsection{Full theory case}\label{AppAsymptoticsFull}

In this section, we will specify the asymptotic behavior of the fields used in Section \ref{sec:FullTheory}. It amounts to a translation of the conditions we are using into the variables we work with.

We start by rephrasing the condition \eqref{AsymptoticsOfMetric} in spherical coordinates such that $x^i = \rho n^i(\theta)$ (note that those coordinates differ from the ones we want to use eventually by a rescaling of the radial coordinate). It reads, in particular,
\begin{subequations}
\begin{align}
	q_{\rho\rho} &= 1 + \frac{1}{\rho}n^in^js_{ij} + O^\infty(\rho^{-1-\epsilon}),\\
	q_{AB} &= \rho^2\eta_{AB} + \rho n^i_{,A}n^j_{,B}s_{ij} + O^\infty(\rho^{1-\epsilon}),
\end{align}
\end{subequations}
where $\eta_{AB}$ is the metric of a unit sphere. To switch to the variables we want to use, we need to find $r$ such that $q_{rr}=1$. In order to do that, we compare lengths of intervals of radial geodesics and obtain the relation
\be
	\int 1\cdot dr = \int\sqrt{q_{\rho\rho}}d\rho,
\ee
which leads to
\be\label{RandRhoRelation}
	r = \rho + \frac{1}{2}n^in^js_{ij}\log\rho + c + O^\infty(\rho^{-\epsilon}).
\ee
Like in the previous section, the role of the $c$ is just to fix the relative values of $r$ and $\rho$, so we choose it to be zero in what follows (it can be restored easily). Moreover, since the angular variables coincide, we can use $n^I$ instead of $n^i$ from now on. Inversion of \eqref{RandRhoRelation} gives
\be
	\rho = r - \frac{1}{2}n^In^Js_{IJ}\log r + O^\infty(r^{-\epsilon}).
\ee
The asymptotics of the metric can now be given explicitly
\begin{subequations}
\begin{align}
	q_{ra} &= \delta_{ra},\\
	q_{AB} &= (r^2 - n^In^Js_{IJ}r\log r)\eta_{AB} + rn^I_{,A}n^J_{,B}s_{IJ} + O^\infty(r^{1-\epsilon}).
\end{align}
\end{subequations}

Transforming the momentum is more involved, because it is a tensor density. Using the notation $\Omega(\theta):=\frac{1}{r^2}\det(\frac{\partial x^I}{\partial y^a})$ and $f^A_I(\theta):=r\frac{\partial y^A}{\partial x^I}$, we find
\begin{subequations}
\begin{align}
	p^r{}_r &= \Omega(\theta)n_I(\theta)n^J(\theta)t^I{}_J(\theta) + O^\infty(r^{-\epsilon}),\\
	p^r{}_A &= r\Omega(\theta)n_I(\theta)n^J_{,A}(\theta)t^I{}_J(\theta) + O^\infty(r^{1-\epsilon}),\\
	p^A{}_B &= \Omega(\theta)f^A_I(\theta)n^J_{,B}(\theta)t^I{}_J(\theta) + O^\infty(r^{-\epsilon}).
\end{align}
\end{subequations}
The fact that the leading terms in the last and second to last lines are not vanishing for large $r$ is the source of the problems with the ``measuring from zero'' implementation of the radial gauge discussed in Section \ref{WhereShouldWeFull}. In the ``measuring from infinity'' implementation, however, such conditions render all the integrals well-defined.

Having the asymptotics of the metric at hand, we can compute the ADM energy. We obtain
\begin{subequations}
\begin{align}
	E &= -\frac{1}{8\pi}\underset{r\rightarrow\infty}{\lim}\left(\int {}^2\!K\sqrt{\det(q_{AB})}d^2\theta - \int {}^2\!\mathring{K}\sqrt{\det(\mathring{q}_{AB})}d^2\theta \right) \\
	   &= \frac{1}{8\pi}\int\left(n^In^J - \frac{1}{2}\eta^{AB}n^I_{,A}n^J_{,B}\right)s_{IJ}\det(\eta)d^2\theta,
\end{align}
\end{subequations}
where in the first line ${}^2\!K$ is (the trace of) the extrinsic curvature of the surfaces of constant $r$ as embedded in the spatial slice $\Sigma$, while the $\mathring{}$ symbol denotes a flat contribution which needs to be subtracted for the final result to be finite \cite{York, GibbonsHawking} (see also \cite{HawkingHorowitz} for a discussion in similar coordinates).

\section{Generalization of the radial gauge to spherically symmetric spacetimes singular at zero}\label{AppSchwarzschild}

Although the ``measuring from infinity'' variant of the reduced phase space construction with the asymptotical behavior of the fields described above seems well suited for a treatment of Schwarzschild black holes, in fact, it is somewhat problematic. The Schwarzschild solution is expressed in variables of \cite{KucharGeometrodynamicsOfThe} by\footnote{Note that in equation (58) of \cite{KucharGeometrodynamicsOfThe}, only the leading orders of the Schwarzschild solution are given. Hence, they satisfy the Hamiltonian constraint only in the limit of large $\rho$ (or $|r|$ in the notation of \cite{KucharGeometrodynamicsOfThe}).}
\be
	\Lambda(\rho) = \frac{1}{\sqrt{1 - \frac{2m}{\rho}}}, \qquad R(\rho) = \rho, \qquad P_\Lambda(\rho) = 0 = P_R(\rho).
\ee
To find expressions for the canonical fields in the radial gauge, we use again the condition \eqref{AsymptoticConditionLambdaProper}. Integrating it (for concreteness we choose $r$ to measure the proper distance from the horizon), we find
\be
	r(\rho) = \rho\sqrt{1 - \frac{2m}{\rho}} + m \log\left(\frac{\rho}{m} - 1 + \frac{\rho}{m}\sqrt{1 - \frac{2m}{\rho}}\right).
\ee
To identify the Schwarzschild solution we should now invert this function, finding $\rho(r)$. Then
\be
	\Lambda(r) = 1, \qquad R(r) = \rho(r),
\ee
would be the Schwarzschild solution in the radial gauge. Unfortunately, inverting that function explicitly is very hard.\footnote{The problem can be reduced to inverting the function $x=(\sqrt{1+y^2}+y)e^y$ for $y(x)$, which is a modified defining problem of the Lambert W function (see \cite{CorlessGonnet}).} It can be checked that the inverse function fulfills the asymptotic behavior spelled out in \eqref{AsymptoticConditionsROfR}, but using it to describe the Schwarzschild solution is highly impractical. Therefore, it is desirable to modify the construction slightly in order to accommodate the Schwarzschild solution more easily.

Being guided by that aim let us recall that Schwarzschild metric expressed in Gullstrand-Painlev\'e coordinates has the form
\be\label{AppSchwarzschildMetric}
	ds^2 = -(1-\frac{2M}{r})dt^2 + 2\sqrt{\frac{2M}{r}}dtdr + dr^2 + r^2 d\Omega^2.
\ee
The idea of using those coordinates in the context of the canonical analysis of spherically symmetric general relativity is not new (see, e.g., \cite{FriedmanLouko,GuvenMurchadha,HusainWinkler} and references therein). They are particularly useful for us, since we see that on $t=\text{const}$ surfaces the variable $r$ measures the proper (spatial) distance. Note however, that the shift vector read off from the above form of the metric is
\be\label{ShiftFromSchwarzAppSchwarz}
	N^r(r) = \sqrt{\frac{2M}{r}},
\ee
which means it does not vanish as $r$ goes to zero, contrary to the case for regular spacetimes considered in Section \ref{SectionSphericalSymmetry}. It does vanish, however, when $r$ goes to infinity. It is this behavior of the shift that we will incorporate in the present modification of the radial gauge construction. Because of this behavior we will work in a setting similar to the ``measuring from infinity'' variant presented in Section \ref{SectionGaugeUnFixingSSFromInf}.

From the form of the metric presented in \eqref{AppSchwarzschildMetric}, we can see that for the Schwarzschild spacetime
\begin{align}\label{AppSchwarzschildFieldsBegin}
	\Lambda(r) = 1,\quad & \quad \dot{\Lambda}(r) = 0,\\
	R(r) = r, \quad & \quad \dot{R}(r) = 0, \\
	N(r) = 1, \quad & \quad N^r(r) = \sqrt{\frac{2M}{r}}.
\end{align}
Using the definitions of canonical momenta spelled out in \eqref{DefMomentaSphericalSymmetry} (with the convention $\chi=1$), we obtain
\be\label{AppSchwarzschildFieldsEnd}
	P_R(r) = \frac{1}{2}\sqrt{\frac{2M}{r}},\quad\quad P_\Lambda(r) = \sqrt{2Mr}.
\ee

Inspecting equation \eqref{eq:SphericalHamiltonianConstraintFromInf}, we see that the reduced Hamiltonian constraint vanishes for the above data. The shift vector preserving the gauge expressed in \eqref{ShiftPreservingGaugeFromInf} is equal to the one read off from the Schwarzschild metric, as written down in \eqref{ShiftFromSchwarzAppSchwarz}. Finally, computing the equations of motion (with the $\chi=1$ convention)
\begin{subequations}\label{EOMsAppSchwarz}
\begin{align}
	\frac{1}{N}\dot{R}(r) &= -\frac{F(r)}{R(r)} - R'(r)\int_r^\infty d\bar{r}\left(\frac{P_R(\bar{r})}{R(\bar{r})} - \frac{F(\bar{r})}{R^2(\bar{r})}\right),\\
	\frac{1}{N}\dot{P}_R(r) &= - R''(r) +\frac{P^2_R(r)}{R(r)} - 2\frac{P_R(r)F(r)}{R^2(r)} + \frac{F^2(r)}{R^3(r)} - P_R'(r)\int_r^\infty d\bar{r} \left(\frac{P_R(\bar{r})}{R(\bar{r})} - \frac{F(\bar{r})}{R^2(\bar{r})}\right),
\end{align}
\end{subequations}
one can verify they are satisfied, leading to the conclusion that a Schwarzschild spacetime is indeed a solution of the equations of motion of the reduced phase space presented in this appendix.

An apparent problem, which arises in this treatment, is that the fields in equations \eqref{AppSchwarzschildFieldsBegin} to \eqref{AppSchwarzschildFieldsEnd} do not satisfy the falloff conditions spelled out in Appendix \ref{AppAsymptotics}. This change of asymptotics of the fields can be understood as stemming from a different choice of slicing in the corresponding spacetime. The falloff behavior of the fields in the current context can be obtained by inspecting what sort of variations of the canonical variables are allowed in order to maintain well-definiteness and functional differentiability of the constraints. It yields the asymptotic behavior which needs to be employed in the phase space presented in this appendix (and propagating in a clear-cut way to the asymtpotic behavior of the corresponding reduced phase space), namely,
\begin{subequations}\label{AppSchwarzschildAsymptoticConditionsR}
\begin{align}
	\Lambda(r) &= 1,\\
	R(r) &= r + O^\infty(r^{-\epsilon}),\\
	P_\Lambda(r) &= \sqrt{2mr} + O^\infty(r^{\frac{1}{2}-\epsilon}),\\
	P_R(r) &= \frac{1}{2}\sqrt{\frac{2m}{r}} + O^\infty(r^{-\frac{1}{2}-\epsilon}),\\
	N(r) &= 1 + O^\infty(r^{-\epsilon}),\\
	N^r(r) &= \sqrt{\frac{2m}{r}} + O^\infty(r^{-\frac{1}{2}-\epsilon}).
\end{align}
\end{subequations}
Additionally, a boundary term equal to the Schwarzschild mass needs to be added to the vector constraint to ensure its functional differentiability, replacing a boundary term for the Hamiltonian constraint present in the previous asymptotic analysis.

\section*{Acknowledgements}

This work was partially supported by the Polish National Science Centre grant No. 2011/02/A/ST2/00300 and by the Polish National Science Centre grant No. 2013/09/N/ST2/04299. NB was supported by a Feodor Lynen Research Fellowship of the Alexander von Humboldt-Foundation.


\bibliographystyle{utphysmendeley}

\end{document}